# Ultrasensitive quantification of dextran sulfate by a mix-and-read fluorescent probe assay

Nathalie Groß[1], Dumitru Arian[1], Ulrich Warttinger[1], Roland Krämer[1]

Correspondence to:

Roland Krämer, phone 0049 6221 548438, fax 0049 6221 548599

E-mail: kraemer@aci.uni-heidelberg.de

1 Heidelberg University, Inorganic Chemistry Institute, Im Neuenheimer Feld 270, 69120 Heidelberg, Germany.

**Abstract**

Dextran sulfate is semi-synthetic, polydisperse sulfated polysaccharide with important applications in clinical practice, in the manufacturing of plasma derived protein therapeutics and in biomedical research. The sensitive detection of dextran sulfate is relevant to preclinical and clinical drug development projects, the quality control of pharmaceutic formulations, and the process control in plasma fractionation using dextran sulfate modified chromatographic columns. Most analytical methods for the sensitive detection of dextran sulfate require multistep protcols and have not been transferred into commercial formats. We describe here the direct quantification of dextran sulfate using the commercially available molecular probe Heparin Red, by a simple mix-and-read fluorescence assay. With a sensitive benchtop fluorimeter, a quantification limit of 310 pg/mL dextran sulfate is achieved. This is superior by at least one order of magnitude to the quantification or detection limit of other reported methods. The outstanding simplicity and sensitivity establish Heparin Red as a new analytical tool for the determination of dextran sulfate.

# Introduction

*Dextran sulfate structure and applications*

Dextran sulfate is semi-synthetic complex polysaccharide. It is prepared by chemical sulfation of dextran, a natural, polydisperse polysaccharide consisiting of a linear backbone of α(1,6) linked glucose monomers (scheme 1, left) that may have branches of smaller chains linked to the backbone by α(1,2), α(1,3) or α(1,4) glycosidic bonds.

Dextran sulfate has important clinical applications. Dextran sulfate–cellulose columns are used as sorbents in blood purification therapies such as removal of low density lipoproteins (LDL apheresis) [1] to treat familial hypercholesterolemia, or of anti-DNA antibodies to treat systemic lupus erythematosus. Low molecular weight dextran sulfate (MW ≈5000 Dalton) has been granted orphan drug designation for mobilization of stem cells from the bone marrow prior to transplantation of hematopoetic stem cells, as well as for pancreatic islet cell transplantation [2]. In experimental nanomedicine, dextran sulfate is a key component of drug delivery systems [3,4]. Dextran sulfate suppresses the effect of amyloid beta on synaptic disfunction and was suggested as a candidate for treatment of Alzheimer's disease [5]. High molecular weight dextran sulfate has inflammogenic properties and is commonly used to induce colitis in mouse models [6]. Dextran sulfate modified chromotagraphic columns are applied in plasma fractionation processes for the isolation of therapeutic plasma proteins, including Factor VIII, factor IX and mannose-binding lectin [7].

*Methods for the quantification of dextran sulfate*

The sensitive detection of dextran sulfate is relevant to preclinical and clinical drug development projects, the quality control of pharmaceutic formulations, and the process control of plasma fractionation, to ensure that products are free of dextrane sulfate traces. Most of the assays described for the quantification of dextran sulfate (examples see table 1) require multistep protocols and/or challenging instrument calibration. Testing with the more recently developed resonance rayleigh scattering and potentiometric membrane methods is relatively simple, although in the latter the user has to prepare the potentiometric sensors for single use.

| Quantification Method | Matrix | Detection limit (ng/mL) | Ref. |
|---|---|---|---|
| Competitive binding assay | plasma | 1000 | [8] |
| Size exclusion HPLC + postcolumn photometry (dimethyl methylene blue) | serum | 300 | [9] |
| Resonance Rayleigh scattering of crystal violet complex | buffer | 13 [a] | [10] |
| Strong ion exchange HPLC + postcolumn photometry (diemthyl methylene blue) | protein solution | 300 | [11] |
| Potentiometric membrane electrode | NaCl solution | 760 | [12] |
| ELISA | buffer | 3 [b] | [13] |

Table 1. Methods for the quantification of dextran sulfate in different matrices, including detection limits.  [a] Triple wavelength analysis; 27 ng/mL for single wavelength analysis.  [b] Given by the provider as lower limit of measureable range for high molecular weight dextran sulfate.

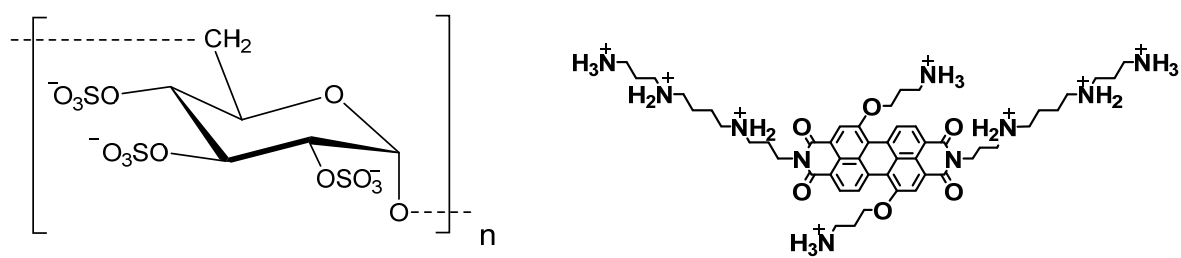

Scheme 1. Left: Repeating 1,6-linked, trisulfated glucose moiety of dextran sulfate. Note that the sulfation degree of DS is variable; see table 1 for the sulfation degree of DS applied in this study. Right: Structure of the polycationic fluorescent probe Heparin Red (right).

*Quantification of sulfated polysaccharides with the fluorescent probe Heparin Red*

We describe here the application of the fluorescent probe "Heparin Red" to the determination of dextran sulfate. Heparin Red has been initially described as an experimental probe for heparin detection [14] and meanwhile developed further into commercially available assays [15] for the detection of sulfated polysaccharides in various matrices, including unfractioned and low-molecular weight heparins [16, 17], chemically modified heparins [16], heparan sulfate [18] and fucoidans [19]. Heparin Red is a polyamine derivative of a red-emissive perylene diimide fluorophore (scheme 1, right). It forms a supramolecular complex with the target, with aggregation of the probe molecules at the heparin template and contact quenching of fluorescence (scheme 2). The strong binding of the polycationic probe to polyanionic heparin appears to be controlled by both electrostatic and aromatic pi-stacking interactions [20].

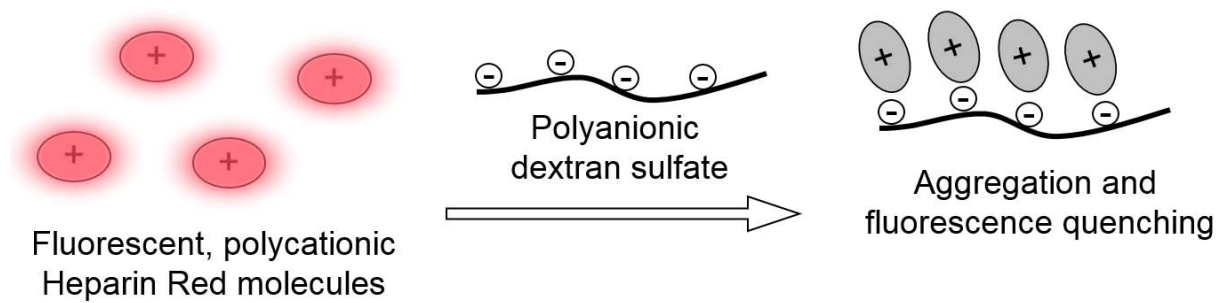

**Scheme 2.** Schematic representation of fluorescence quenching of the molecular probe Heparin Red in the presence of dextran sulfate.

In this contribution, we describe for the first time the application of the commercial reagent Heparin Red to the highly sensitive, direct quantification of dextran sulfate by a simple mix-and-read cuvette assay.

## Materials and Methods

**Instrumentation**

*Fluorescence measurements*
Fluorescence was measured with the benchtop fluorimeter FluoroLog®-3 (Horiba Scientific, Kyoto, Japan), excitation at 570 nm, emission recorded at 590-650 nm, integration time 0.4 sec, average of 3 scans, spectral band width Ex/Em 7nm.

*Pipettes*
Transferpette® 0,5-10µl, purchased from Brand GmbH, Wertheim. Rainin Pipettes 100-1000µl, 20-200µl purchased from Mettler Toledo, OH, USA, Eppendorf Multipette M4, 50mL, purchased from VWR international, PA, USA.

**Reagents**

*Heparin Red®*

Heparin Red® solution 100 µM, as a component of the Heparin Red® Kit, was a gift from Redprobes UG, Münster, Germany, www.redprobes.com . Product No HRU001, Lot 003.

*Dextran sulfate*

Dextran sulfate sodium salt, product number 42867, Lot # BCBP2520V, was purchased from Sigma Aldrich GmbH, Steinheim, Germany. Sulfur content 16,15% according to certificate of analysis. CHN analysis performed in our laboratory (average of two analyses): C 17,18 %, H 3,54 % , N 1,55 %. N-content of the sample may arise from residual nitrogen-containg reagents (chlorosulfonic acid-pyridine or amidosulfonic acid), commonly used [21] for the chemical sulfation of dextran. No precipitate or cloudiness is observed after addition of $Ba(NO_3)_2$ to a 20 mg/mL solution of the dextran sulfate, indicating the absence of significant levels of free sulfate ($SO_4^{2-}$).

*Other*

All aqueous solutions were prepared with HPLC grade water purchased from VWR, product No 23595.328. Dimethyl sulfoxide (DMSO), product number 34869, Lot # STBF6384V, and Hydrochloric acid (1,0 M), product number 35328, Lot # SZBF2050V, were purchased from Sigma Aldrich GmbH, Steinheim.

**Assays**

*Heparin Red® Assay for dextran sulfate*

A 6,0 nM Heparin Red solution in DMSO containing 30 mM HCl was freshly prepared by mixing appropriate volumes of DMSO, 1,0 M HCl and 100 µM Heparin Red solution. This mixture was stored at ambient temperature and used within 7 hours. Aqueous dextran sulfate solutions at 210 ng/mL (for titration and calibration curve) and 420 pg/mL (spiked sample) were freshly prepared from a 2,1 mg/mL dextran sulfate stock solution and used within 2 hours. 2,5 mL of the aqueous sample were placed in a 4,5 mL polystyrene fluorescence cuvette, 0,5 mL of the Heparin Red solution added, the cuvette sealed with a cap and briefly mixed by vortexing. The cuvette was then placed into the fluorimeter and fluorescence (590 – 650 nm) recorded within 2 minutes (fluorescence of the mixture is stable for at least 20 minutes). The calibration curve (figure 1) was derived by mixing 2,5 mL water and 0,5 mL

Heparin Red solution as described above and titrating it with 5 µL aliquots of the 210 ng/mL dextran sulfate solution (vortexing after each addition).

## Results and discussion

### Detection of dextran sulfate in aqueous matrix

Selected properties of the dextran sulfate sample applied in this study are listed in table 2.

|  | Average mol. Weight [a] | Average sulfation (per monosaccharide) [b] | Average charge density (per monosachharide) |
|---|---|---|---|
| Dextran sulfate | 40 kD | 2,1 | -2,1 |

Table 2. Selected properties of the commercial dextran sulfate sodium salt used in this study. a) Taken from the certificate of analysis of the provider. b) Calculated using the ratio between S-content (16,15 % according to certificate of analysis of the provider) and C-content (17,18 % by microanalysis in our laboratory) of the dextran sulfate sample. This approach for the sulfation degree implies that the sample is free of C- or S-containing impurities.

A very sensitive benchtop fluorimeter was used for fluorescence measurements to ensure good signal-to-noise ratio at the low dye concentration applied. A 6,0 nM solution of Heparin Red in DMSO/30 mM HCl was prepared and applied to all determinations shown in figure 1. The calibration curve was derived by "titrating" a mixture of 0,5 mL of this Heparin Red solution and 2,5 mL water with dextran sulfate, leading to the expected quenching of dye fluorescence with increasing dextran sulfate concentration. A linear response is observed in the concentration range 0 – 2,1 ng/mL. Accurate recovery of a dextran sulfate spike was verified by mixing 2,5 mL of a 420 pg/mL aqueous solution with 0,5 mL Heparin Red solution. Fluorescence ($F_0$) and standard deviation of the blank, i.e. an aqueous sample without dextran sulfate, was obtained by measuring mixtures of 2,5 mL water and 0,5 mL Heparin Red solution, each (n=6) in a different cuvette.

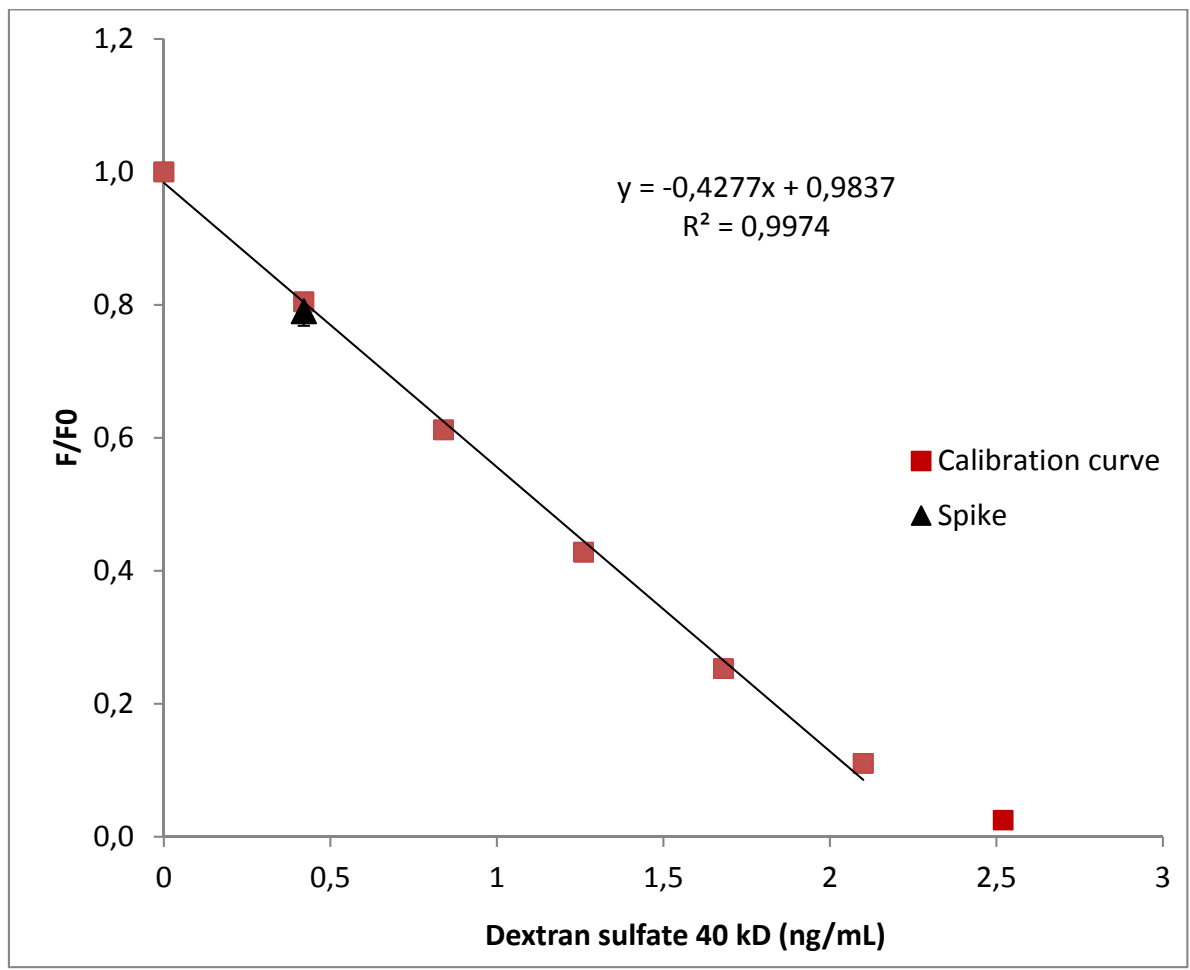

**Figure 1.** Fluorescence response of Heparin Red to dextran sulfate. Excitation at 570 nm, fluorescence emission recorded at 612 nm. Solutions were allowed to equilibrate to ambient temperature (20°C) before mixing. Fluorescence corrected for the background value of pure water. F0 is the average fluorescence of 6 blanks (CV = 0,7 %), i.e. mixtures of 2,5 mL water and 0,5 mL Heparin Red solution (6 nM in DMSO/30 mM HCl) in polystyerene cuvettes. Calibration curve: 2,5 mL water and 0,5 mL Heparin Red solution (6 nM in DMSO/30 mM HCl) were mixed in a polystyrene cuvette and the mixture "titrated" with 5 µL aliquots of a 210 ng/mL dextran sulfate solution. This titration was performed twice in different cuvettes, average F/F0 values are given (CV averaged over all concentrations within linear range: 1,3 %). The indicated dextran sulfate concentration does *not* correspond to the actual concentration in the reaction mixture (3 mL) but is, for better comparison with the spike, related to the 2,5 mL aqueous portion. Spike: 2,5 mL aqueous dextran sulfate (420 pg/mL) and 0,5 mL Heparin Red solution (6 nM in DMSO/30 mM HCl) were mixed in a polystyrene cuvette; average of triplicate determination, CV = 2,7%.

By extrapolation of the linear concentration range in figure 1, a theoretical intersection point with the fluorescence baseline at excess dextran sulfate is derived at 2,3 ng/mL dextran sulfate. This is somewhat higher than the calculated value 1,9 ng/mL for the formation of charge neutral, fluorescence-quenched aggregates with Heparin Red [20], considering the charge +8 for the fully protonated probe (scheme 1) and 5 pMol "negative charge" per ng dextran sulfate, as derived from the S-content 16,15% of the dextran sulfate sample (assuming that each S atom contributes one $-SO_3^-$ residue). The linear decrease of fluorescence indicates a smooth formation of the aggregate even at very low concentrations of the components.

**Quantification limit**

The quantification limit of the method (LOQ) was determined based on signal-to-noise [22], by comparing measured signals from samples with known low concentrations of analyte with those of blank samples. The LOQ was derived using the equation

$$x_Q = \bar{x}_{blank} - 10\, \sigma_{blank}$$

where $x_Q$, is the smallest measure that can be quantified with reasonable certainty for the given analytical procedure, $\bar{x}_{blank}$ is the mean of the blank measures, and $\sigma_{blank}$ is the standard deviation of the blank measures.

For the series of fluorescence measurements included in figure 1, $\bar{x}_{blank}$ = 1 and $\sigma_{blank}$ = 0,007 (n=6). A more representative value for $\sigma_{blank}$ is 0,012, obtained as an average of 4 independent detections (n between 5 and 8), from which the highest $\sigma_{blank}$ was 0,015 . Using the latter value, $x_Q$ = 0,85 is calculated as the smallest quantitative fluorescence measure, what corresponds to a dextran sulfate LOQ = 310 pg/mL when compared with the linear range of the calibration curve. Recovery of a 420 pg/mL dextran sulfate spike with good precision (figure 1) confirms that pg/mL levels can indeed by quantified by the method. A refined analysis of spike recovery at the quantification and detection limit and the investigation of matrix effects in progress.

## Conclusion

The sensitive quantification of dextran sulfate, a semi-synthetic sulfated polysaccharide, is relevant to clinical practice, the manufacturing of plasma derived protein therapeutics and to biomedical research. Current analytical methods often require multistep protocols and have quantification or detection limits in the ng/mL range. We describe here a simple mix-and-read fluorescence assay for the determination of dextran sulfate with a quantification limit in the pg/mL range. The outstanding simplicity and sensitivity of the assay establish the fluorescent probe Heparin Red as a new tool for the quantification of dextran sulfate.

**Conflict of interest.** R.K. holds shares in Redprobes UG, Münster, Germany. Other authors: No conflict of interest.